\documentclass[%
 reprint,
superscriptaddress,
 amsmath,amssymb,
 aps,
prc,
]{revtex4-1}

\usepackage{color}
\usepackage{graphicx}
\usepackage{dcolumn}
\usepackage{bm}
\usepackage[mathlines]{lineno}

\begin{document}

\preprint{APS/123-QED}

\title{Search for Two-Neutrino Double Electron Capture of $^{124}$Xe with XENON100}

\newcommand{\bern}{\affiliation{Albert Einstein Center for Fundamental Physics, University of Bern, Bern, Switzerland}}
\newcommand{\bologna}{\affiliation{Department of Physics and Astrophysics, University of Bologna and INFN-Bologna, Bologna, Italy}}
\newcommand{\chicago}{\affiliation{Department of Physics \& Kavli Institute of Cosmological Physics, University of Chicago, Chicago, IL, USA}}
\newcommand{\coimbra}{\affiliation{Department of Physics, University of Coimbra, Coimbra, Portugal}}
\newcommand{\columbia}{\affiliation{Physics Department, Columbia University, New York, NY, USA}}
\newcommand{\lngs}{\affiliation{INFN-Laboratori Nazionali del Gran Sasso and Gran Sasso Science Institute, L'Aquila, Italy}}
\newcommand{\mainz}{\affiliation{Institut f\"ur Physik \& Exzellenzcluster PRISMA, Johannes Gutenberg-Universit\"at Mainz, Mainz, Germany}}
\newcommand{\heidelberg}{\affiliation{Max-Planck-Institut f\"ur Kernphysik, Heidelberg, Germany}}
\newcommand{\munster}{\affiliation{Institut f\"ur Kernphysik, Wilhelms-Universit\"at M\"unster, M\"unster, Germany}}
\newcommand{\nikhef}{\affiliation{Nikhef and the University of Amsterdam, Science Park, Amsterdam, Netherlands}}
\newcommand{\nyuad}{\affiliation{New York University Abu Dhabi, Abu Dhabi, United Arab Emirates}}
\newcommand{\purdue}{\affiliation{Department of Physics and Astronomy, Purdue University, West Lafayette, IN, USA}}
\newcommand{\rpi}{\affiliation{Department of Physics, Applied Physics and Astronomy, Rensselaer Polytechnic Institute, Troy, NY, USA}}
\newcommand{\rice}{\affiliation{Department of Physics and Astronomy, Rice University, Houston, TX, USA}}
\newcommand{\stockholm}{\affiliation{Oskar Klein Centre, Department of Physics, Stockholm University, AlbaNova, SE-106 91 Stockholm, Sweden}}
\newcommand{\subatech}{\affiliation{SUBATECH, Ecole des Mines de Nantes, CNRS/In2p3, Universit\'e de Nantes, Nantes, France}}
\newcommand{\torino}{\affiliation{INFN-Torino and Osservatorio Astrofisico di Torino, Torino, Italy}}
\newcommand{\ucla}{\affiliation{Physics \& Astronomy Department, University of California, Los Angeles, CA, USA}}
\newcommand{\ucsd}{\affiliation{Department of Physics, University of California, San Diego, CA, USA}}
\newcommand{\wis}{\affiliation{Department of Particle Physics and Astrophysics, Weizmann Institute of Science, Rehovot, Israel}}
\newcommand{\zurich}{\affiliation{Physik-Institut, University of Zurich, Zurich, Switzerland}}

\author{E.~Aprile}\columbia
\author{J.~Aalbers}\nikhef
\author{F.~Agostini}\lngs\bologna
\author{M.~Alfonsi}\mainz
\author{F.~D.~Amaro}\coimbra
\author{M.~Anthony}\columbia
\author{F.~Arneodo}\nyuad
\author{P.~Barrow}\zurich
\author{L.~Baudis}\zurich
\author{B.~Bauermeister}\stockholm\mainz
% ...
\collaboration{XENON Collaboration}\email{xenon@lngs.infn.it}\noaffiliation

\date{\today}

\begin{abstract}
Two-neutrino double electron capture is a rare nuclear decay where two electrons are simultaneously captured from the atomic shell. For $^{124}$Xe this process has not yet been observed and its detection would provide a new reference for nuclear matrix element calculations.
We have conducted a search for two-neutrino double electron capture from the K-shell of $^{124}$Xe using 7636\,kg$\cdot$d of data from the XENON100 dark matter detector. Using a Bayesian analysis we observed no significant excess above background, leading to a lower 90\% credibility limit on the half-life $T_{1/2}>6.5\times10^{20}$\,yr. We have also evaluated the sensitivity of the XENON1T experiment, which is currently being commissioned, and found a sensitivity of $T_{1/2}>6.1\times10^{22}$\,yr after an exposure of 2\,t$\cdot$yr.
\end{abstract}

\maketitle


\section{\label{sec:level1}Introduction}

Double electron capture is a rare nuclear decay where a nucleus captures two electrons from the atomic shell
\begin{equation}
(A, Z) + 2e^- \rightarrow (A, Z-2) + (2\nu_e)\,.
\end{equation}
The two-neutrino mode (2$\nu$2EC) is allowed in the Standard Model while the existence of the lepton number-violating neutrinoless double electron capture (0$\nu$2EC) would prove the Majorana nature of the neutrino. In 0$\nu$2EC, there is the possibility of a resonant enhancement of the decay rate for decays to excited atomic or nuclear states \cite{sujkowski03,krivoruchenko10,suhonen13a,maalampi13}. Due to low isotopic abundances and longer half-lives \cite{rodriguez12,kotila14}, experimental searches for 0$\nu$2EC are generally not competitive with those for neutrinoless double beta decay (0$\nu$2$\beta$) to constrain the effective neutrino mass $m_{\beta\beta}$ and the neutrino mass hierarchy \cite{povinec15}.
The largest uncertainty in the conversion of the half-life of 0$\nu$2$\beta$ or 0$\nu$2EC to $m_{\beta\beta}$ is introduced by the calculation of nuclear matrix elements. Although the matrix elements for the two-neutrino and neutrinoless modes of the double electron capture differ, they are based on the same nuclear structure models. A measurement of the 2$\nu$2EC half-life would help to test the accuracy of these models.

So far, 2$\nu$2EC has only been observed for $^{130}$Ba in geochemical experiments \cite{meshik01,pujol09}. In addition, there is an indication for 2$\nu$2EC of $^{78}$Kr from a low-background proportional counter \cite{gavrilyuk13}. 
In natural xenon the isotopes $^{124}$Xe ($Q=2864$\,keV \cite{wang12}, abundance 0.095\% \cite{berglund11}) and $^{126}$Xe ($Q=919$\,keV \cite{wang12}, abundance 0.089\% \cite{berglund11}) can decay via 2$\nu$2EC. However, any signal will be dominated by $^{124}$Xe due to the $Q^5$ dependence of the phase space \cite{bernabeu1983}. In the case of $^{124}$Xe, the theoretically calculated branching ratio that the two electrons are captured from the K-shell (2$\nu$2K) is 76.7\% \cite{gavrilyuk15}. Filling the vacancies of the daughter atom $^{124}$Te leads to the emission of X-rays and Auger electrons with a total energy of approximately 64\,keV. There is a wide spread in the predicted half-lives of 2$\nu$2EC for $^{124}$Xe, from $\sim10^{20}$\,yr to $10^{24}$\,yr due to different nuclear matrix element calculations \cite{hirsch94,aunola96,rumyantsev98,suhonen13b,singh07,shukla07}. Previous searches for 2$\nu$2K of $^{124}$Xe have been carried out using a low-background proportional counter with enriched xenon \cite{gavrilyuk16,gavrilyuk15} and large-scale liquid xenon detectors \cite{mei13,abe15}. The current best experimental limit on the half-life, $T_{1/2}>4.7\times10^{21}$\,yr (90\% confidence level), is set by the XMASS experiment \cite{abe15}.  

A limit of $T_{1/2}>1.66\times10^{21}$\,yr (90\% confidence level) was derived from previously published XENON100 data \cite{mei13}. However, the available data was not well suited for a signal search due to the coarse binning. The limit was calculated from the average background rate for the energy region below $\sim10$\,keV \cite{aprile12a}, outside the expected 2$\nu$2K signal region and the assumed isotopic abundance of $^{124}$Xe did not match the real situation. Here, we improve on this study by using the 224.6 live days of XENON100 data and additional insight into the experimental details.

\section{\label{sec:level2}The XENON100 Experiment}
Located at the Laboratori Nazionali del Gran Sasso (LNGS), the XENON100 experiment \cite{aprile11a} utilizes a dual-phase xenon time-projection chamber (TPC) in order to search for dark matter particles in form of Weakly Interacting Massive Particles (WIMPs). 
The TPC contains a total 62\,kg of liquid xenon (LXe) in a cylindrical (30.5\,cm height and diameter) volume equipped with 178 radio pure photomultiplier tubes (PMTs) placed in the gaseous phase on top and immersed in the LXe at the bottom. The TPC is fully surrounded by an active LXe veto viewed by 64 additional PMTs.
If a particle deposits its energy in the LXe, it creates excited atoms and ions leading to the formation of excimers. The de-excitation of these excimers causes prompt scintillation light ($S1$). A fraction of the electrons generated by the ionization process are drifted towards the gas phase by an applied electric field of 530\,V/cm. At the liquid-gas interface they are extracted and accelerated by a strong field of 12\,kV/cm. This induces a secondary scintillation signal ($S2$) which is proportional to the number of generated electrons. Three-dimensional event vertex reconstruction is achieved by obtaining the interaction depth from the time difference of the two signals and by deriving the $(x,y)$-position from the hit pattern of the $S2$ signal on the top PMT array. A background-optimized fiducial volume can thus be selected, with a strongly reduced background from external $\gamma$-radiation. A detailed description of the detector can be found in Ref.~\cite{aprile11a}.

\section{\label{sec:level3}Data Analysis}
Data for the 2$\nu$2K-search consists of 224.6 live days collected between February 28,\,2011 and March 31,\,2012 using a fiducial target mass of 34\,kg. This data set has also been analyzed for different purposes in Refs.~\cite{aprile12a,aprile13a,aprile14a,aprile15a,aprile15b}.
The detector was filled with a mixture of natural xenon and xenon depleted in $^{136}$Xe and $^{124}$Xe, leading to a $^{124}$Xe abundance of $\eta=(8.40\pm 0.07)\times10^{-4}$. This corresponds to an absolute amount of about 29\,g of $^{124}$Xe in the fiducial volume.

The energy calibration uses the $S1$ and $S2$ signals from an Americium-Beryllium ($^{241}$AmBe) calibration measurement. We employ $\gamma$-lines from neutron-activated xenon isotopes at 40\,keV and 320\,keV ($^{129}$Xe), 80\,keV ($^{131}$Xe), 164\,keV ($^{131m}$Xe) and 236\,keV ($^{129m}$Xe). The energy $E$ was obtained by a linear combination of the $S1$ and $S2$ signals measured in photoelectrons (PE), exploiting their anti-correlation \cite{aprile07,szydagis11}
\begin{equation}
E = W \cdot \left({\frac{S1}{g_1} + \frac{S2}{g_2}}\right)\,.
\end{equation}
$W$ is the mean energy required to produce a photon or electron and $g_1$ and $g_2$ are detector-specific gain factors. When fixing $W$ to 13.7\,eV \cite{szydagis11}, the best-fit values of the gain factors are \mbox{$g_1=(5.07\pm0.03)\times10^{-2}$\,PE/photon} and \mbox{$g_2=(10.13\pm0.07)$\,PE/electron}. Although the $S1$ and $S2$ signals depend on particle type and vary non-linearly with energy, the combined signal provides a common, linear energy scale for both X-rays and Auger electrons at the relevant energies.
The energy resolution $\sigma$ was derived from the same \mbox{$\gamma$-lines} and is given by
\begin{equation}
\sigma(E)=a \cdot \sqrt{E}+ b \cdot E\,,
\label{eq:sigma}
\end{equation}
with $a=(0.405\pm0.010)\,\sqrt{\mathrm{keV}}$ and $b=0.0261\pm0.0008$.

Selection cuts were applied in order to ensure data quality and consistency. Every valid event was required to have exactly one S1 and one corresponding S2. In order to avoid dark count contributions, the S1 had to be detected by at least two PMTs. Further cuts address the removal of noisy events using the information on the signal width ($S1$ and $S2$) and on the signal distribution between the top and bottom PMT arrays. Events which have a coincident signal in the veto were not considered as this indicates multiple scattering induced by external radiation. The acceptance of each selection cut was calculated analogous to Ref.~\cite{aprile14b} by determining the fraction of events that passed all selection cuts but the one of interest. The total acceptance $\epsilon$ in the $\pm3\sigma$-region around the expected signal was found to be $\epsilon=(98.3\pm0.1)\%$. 

\begin{figure}
 \centering
  \includegraphics[width = 0.5\textwidth]{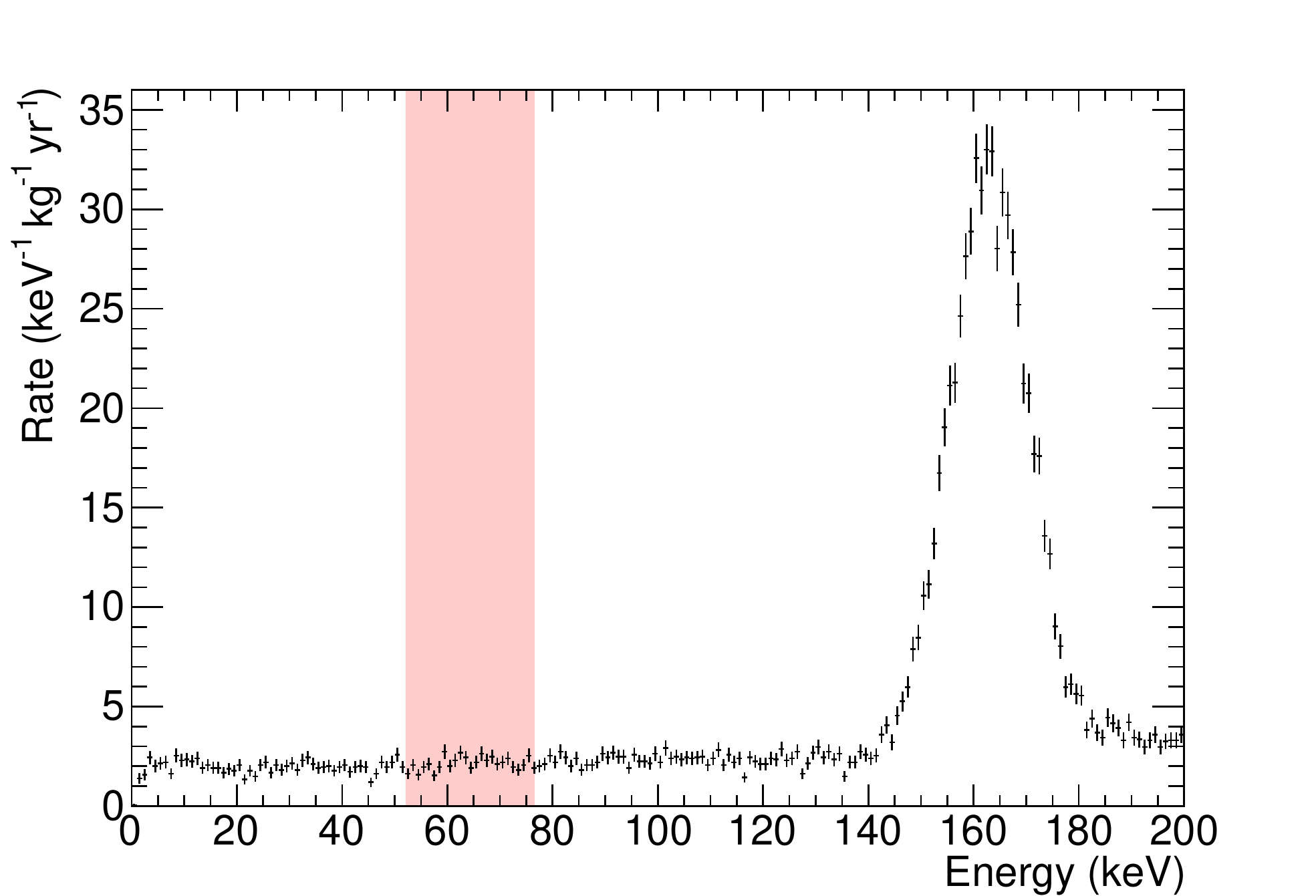}
\caption{\label{fig:spectrum} Spectrum of remaining events after all cuts. The peak at 164\,keV originates from $^{131m}$Xe. The red shaded area indicates the $\pm3\sigma$ region around the expected 2$\nu$2K peak of $^{124}$Xe.}
 \end{figure}

Fig.~\ref{fig:spectrum} shows the spectrum after applying all cuts. The peak at 164\,keV originates from the de-excitation of the long-lived $^{131m}$Xe ($\tau=11.8$\,d) from the neutron calibration before the run. Apart from this peak, the background is nearly constant with an average rate of $5.9\times10^{-3}$\,events/(keV\,$\cdot$\,kg\,$\cdot$\,day). This is expected for a background that is dominated by low-energy Compton scatters \cite{aprile11b}.

To determine the expected mean energy and width of the signal, we calculated the energies and emission probabilities of all X-rays and Auger electrons for a single K-shell vacancy in Te using the RELAX code \cite{relax}. The calculation accounts for bound-bound X-rays and electrons which are emitted from transitions within the atomic shell. In addition, it is assumed that the final atom returns to neutrality by filling all remaining vacancies with electrons from the continuum which leads to the emission of free-bound X-rays. The results are summarized in Table~\ref{tab:x-rays}.

\begin{table}
\caption{\label{tab:x-rays}Average energies and average number of emitted quanta per K-shell vacancy in Te, as calculated with the RELAX code \cite{relax}.}
\begin{ruledtabular}
\begin{tabular}{lrd}
 & \multicolumn{1}{c}{\textrm{Energy per}} &  \multicolumn{1}{c}{\textrm{Number of}} \\
 &  \multicolumn{1}{c}{\textrm{Quantum (eV)}} &  \multicolumn{1}{c}{\textrm{Quanta}} \\
\hline
bound-bound X-rays & 25950 & 0.96 \\
bound-bound electrons & 572 & 11.7 \\
free-bound X-rays & 14 & 12.7 \\
\end{tabular}
\end{ruledtabular}
\end{table}

The individual quanta emitted in the de-excitation process cannot be resolved due to the limited spatial and timing resolution of the detector. Therefore, the expected signal is a single peak at the sum energy.
The RELAX code assumes that the shell binding energies are independent of the ionization of the atom. Therefore, the total emitted energy equals the binding energy of the K-shell of the neutral atom $E_K=31.8$\,keV \cite{xrdbl}.
In 2$\nu$2K the total emitted energy is given by the double-electron hole energy $(64.457\pm0.012)$\,keV \cite{nesterenko2012} which is very close to two times the K-shell binding energy $2E_K=63.6$\,keV.
However, the energies of a small fraction of the emitted quanta are below the xenon excitation threshold of 13.7\,eV. According to  RELAX, and under the assumption that the quanta emitted in 2$\nu$2K are similar to those generated by two single K-shell vacancies, this leads to an average energy loss of 0.13\,keV. Therefore, the 2$\nu$2K peak is expected to be centered at 64.33\,keV.
To estimate the energy resolution of the signal peak, we take the average energy and number of X-rays and electrons emitted per vacancy as shown in Table~\ref{tab:x-rays} but neglect the contribution from free-bound X-rays. Again, assuming the same de-excitation spectrum as for two single K-captures, we arrive at an energy resolution $\sigma_\mathrm{sig}$ of  
\begin{equation}
\sigma_\mathrm{sig}=\sqrt{2\cdot(n_1\cdot\sigma_1^2+n_2\cdot\sigma_2^2)}~,
\end{equation}
where $n_1=0.96$ and $n_2=11.7$ are the average number of X-rays and Auger electrons and $\sigma_1=\sigma(25.9\,\mathrm{keV})$ and $\sigma_2=\sigma(0.57\,\mathrm{keV})$ correspond to their respective energy resolution from Eq.~(\ref{eq:sigma}). This leads to \mbox{$\sigma_\mathrm{sig}=(4.10\pm0.27)$\,keV} compared to \mbox{$\sigma(64.33\,keV)=(4.93\pm0.27)$\,keV} for a single energy deposition.

The statistical analysis for the signal search uses the Bayesian Analysis Toolkit \cite{caldwell08}.
The spectrum with a 1\,keV binning was fit in the energy range \mbox{10--135}\,keV with a ``signal+background'' model $f_\mathrm{sig}(E)$ and a ``background-only'' model $f_\mathrm{bkg}(E)$
\begin{eqnarray}
\label{eq:fit_functions}
f_\mathrm{sig}(E)&=& \frac{\Gamma \epsilon \eta  mt N_A}{\sqrt{2\pi} \sigma_\mathrm{sig} M_\mathrm{Xe}}\exp{\left(-\frac{(E-\mu_\mathrm{sig})^2}{2\sigma_\mathrm{sig}^2}\right)}\nonumber\\
&&+f_\mathrm{bkg}(E)\,,\\
f_\mathrm{bkg}(E) &=& a_\mathrm{bkg}E+c_\mathrm{bkg}\,.
\end{eqnarray}
$E$ is the energy, $\Gamma$ the decay rate, $\epsilon$ the signal acceptance, $\eta$ the abundance of $^{124}$Xe, $mt$ the exposure, $N_A$ Avogadro's constant, $M_\text{Xe}$ the molar mass of xenon, $\mu_\mathrm{sig}$ is the mean energy and $\sigma_\mathrm{sig}$ the width of the signal peak. The parameters $a_\mathrm{bkg}$ and $c_\mathrm{bkg}$ represent the slope and constant term of the background spectrum, respectively.
The binned likelihood of the fit assumes independent Poisson fluctuations of the bin entries and is defined as
\begin{equation}
\mathcal{L} = \prod_{i=1}^{N_\mathrm{bin}}\frac{\lambda_i^{n_i}e^{-\lambda_i}}{n_i!}\,,
\end{equation}
where $N_\mathrm{bin}$ is the total number of bins. $\lambda_i$ is the expected number of events and $n_i$ the observed number of events in the $i$th bin.
Systematic uncertainties were included in the fit by Gaussian priors and are summarized in Table~\ref{tab:parameters}. The uncertainty on the cut acceptance $\epsilon$ is only statistical. For the natural abundance $\eta$, the uncertainty was calculated from the individual uncertainties on the amounts and abundances of the deployed xenon batches.
The uncertainty on the exposure $mt$ accounts for the uncertainty in the determination of the fiducial volume due to the limited spatial resolution. Regarding the peak position $\mu_\mathrm{sig}$ and width $\sigma_\mathrm{sig}$, we included the uncertainties derived from the fits to the energy calibration and resolution. In addition, we added systematic uncertainties of 0.2\% and 3\% for the peak position and energy resolution, respectively, which were determined from the RELAX calculation.
Uniform priors were chosen for the remaining free parameters of the fit, $\Gamma$, $a_\mathrm{bkg}$ and $c_\mathrm{bkg}$.
All fit parameters were constrained to physically allowed positive values.
The significance of a possible signal was evaluated by calculating the Bayes Factor \cite{kass95}
\begin{equation}
B=\frac{P(f_\mathrm{bkg} \mid \vec D)}{P(f_\mathrm{sig} \mid \vec D)}\,,
\end{equation}
where $P(f \mid \vec D )$ is the posterior probability of the model $f$ and $\vec D$ is the data.

\begin{table}
\caption{\label{tab:parameters}Gaussian priors included in the fit to account for systematic uncertainties. The value and uncertainty denote the mean and standard deviation of the Gaussian prior, respectively.}
\begin{ruledtabular}
\begin{tabular}{lr}
\textrm{Parameter} & \textrm{Value} \\
\hline
acceptance $\epsilon$ & $(98.3\pm0.1)\%$ \\
abundance $\eta$ & $(8.40\pm0.07)\times10^{-4}$ \\
exposure $mt$ & $(7636\pm45)$\,kg$\cdot$d \\
peak position $\mu_\mathrm{sig}$ & $(64.33\pm0.37)$\,keV \\
peak width $\sigma_\mathrm{sig}$ & $(4.10\pm0.27)$\,keV \\
\end{tabular}
\end{ruledtabular}
\end{table}

\section{\label{sec:level4}Results}
\begin{figure}
 \centering
 \includegraphics[width = 0.5\textwidth]{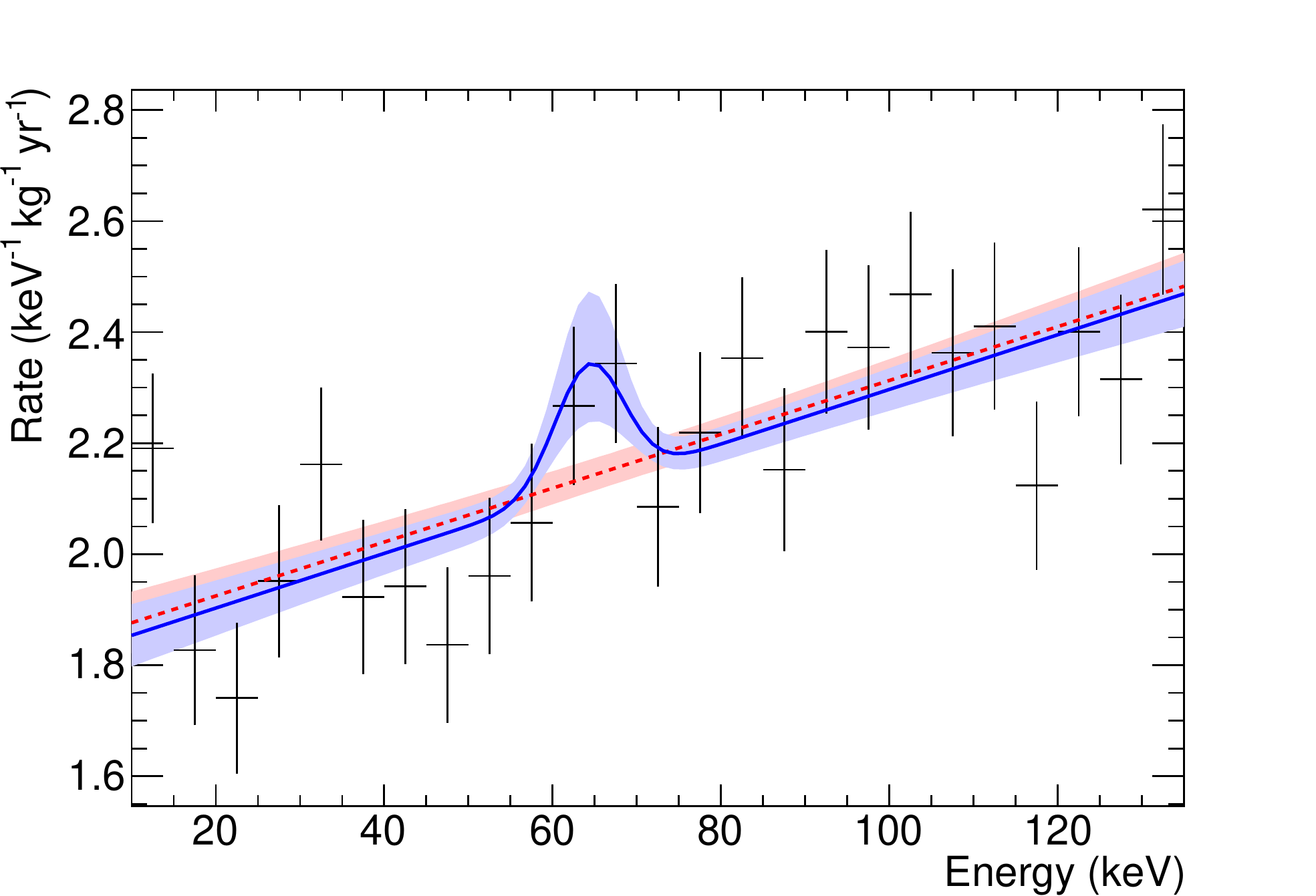}
 \caption{\label{fig:fits} Best fit to the data with the ``signal+background'' model $f_\mathrm{sig}$ (blue solid line) and the ``background-only'' model $f_\mathrm{bkg}$ (red dashed line). The shaded areas indicate the 68\% uncertainty bands.}
 \end{figure}

\begin{figure}
 \centering
 \includegraphics[width = 0.5\textwidth]{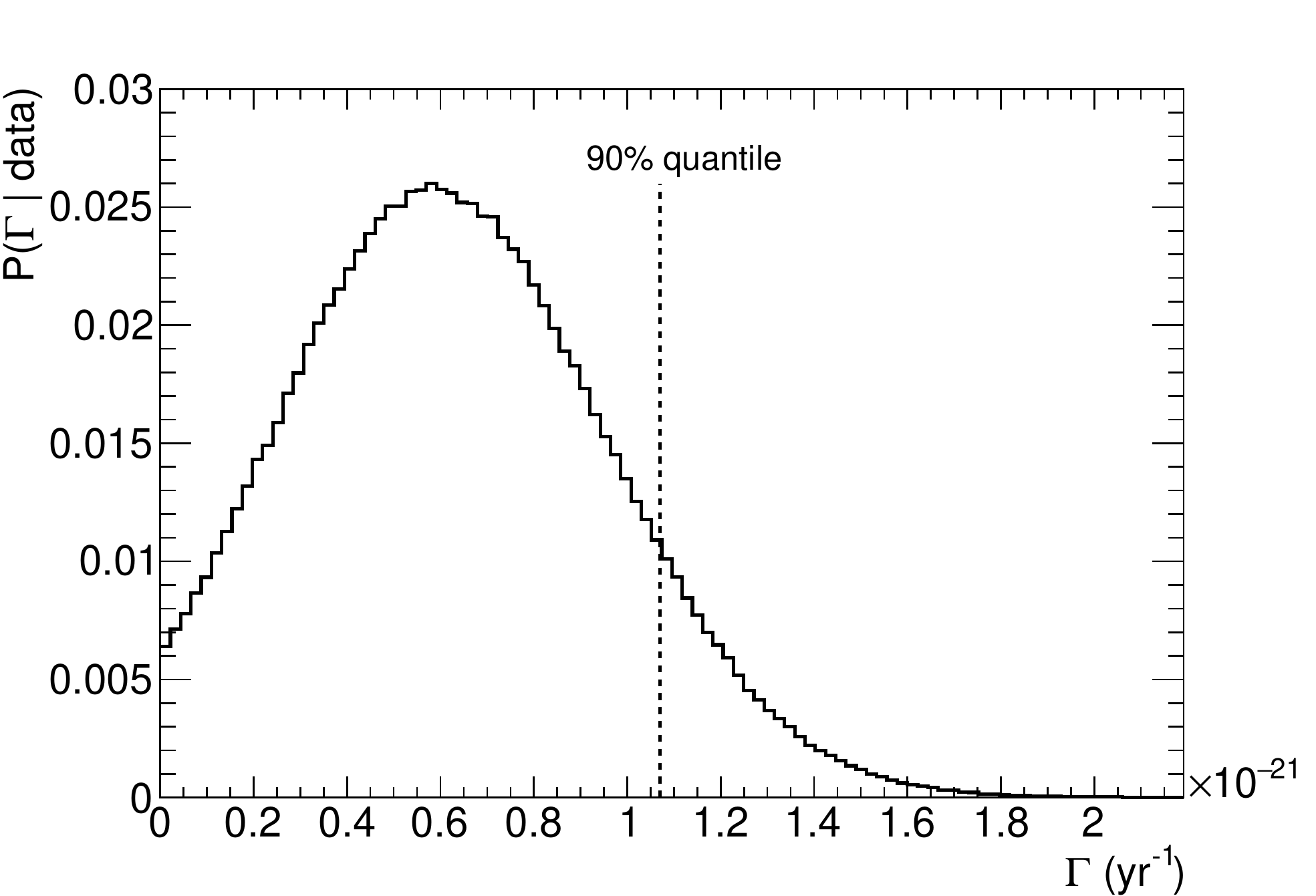}
 \caption{\label{fig:posterior} Marginalized posterior probability distribution for the decay rate $\Gamma$. The vertical line indicates the 90\% quantile from which the lower half-life limit was derived.}
 \end{figure}

The best fits to the spectrum with $f_\mathrm{sig}$ and $f_\mathrm{bkg}$ are shown in Fig.~\ref{fig:fits} and the obtained values can be found in Tab.~\ref{tab:best_fit_parameters}. The p-values of the ``signal+background'' and ``background-only'' fit, calculated as described in Ref.~\cite{beaujean11}, are 0.92 and 0.89, respectively. These values show that the data is well described by both fit models.
Since the Bayes Factor is 
\begin{equation}
B=1.2\,,
\end{equation}
and thus favors the ``background-only'' model, we calculate a lower limit on the half-life.
The 90\% credibility limit $\Gamma_\mathrm{lim}$ on the decay rate is defined as the 90\% quantile of the marginalized posterior probability distribution shown in Fig.~\ref{fig:posterior}.
This leads to a 90\% credibility limit on the half-life $T_{1/2}$ of
\begin{equation}
T_{1/2}>\frac{\ln(2)}{\Gamma_\mathrm{lim}}=6.5\times10^{20}\,\mathrm{yr}\,.
\end{equation}
The influence of the nuisance parameters on the limit was evaluated by fixing all parameters shown in Table~\ref{tab:parameters} to their mean values, which weakens the limit by 0.5\%. While the binning has only a moderate influence ($\sim15\%$), decreasing the fit range can worsen the half-life limit by up to a factor of $\sim2$. The latter comes from the anti-correlation of the parameters $\Gamma$ and $c_\mathrm{bkg}$ in Eq.~(\ref{eq:fit_functions}).
We have checked our result with the Feldman-Cousins procedure \cite{feldman97} using the approach of a simple counting experiment with known background. The number of observed events was derived from the $\pm3\sigma$ region of the expected 2$\nu$2K peak, while the number of background events was calculated from the regions left and right of the peak. This gives a lower half-life limit of $7.3\times10^{20}$\,yr (90\% confidence level) which confirms the result of the full Bayesian analysis. 

\begin{table}
\caption{\label{tab:best_fit_parameters}Best fit parameters obtained for the data with the ``signal+background'' model $f_\mathrm{sig}$ and the ``background-only'' model $f_\mathrm{bkg}$.}
\begin{ruledtabular}
\begin{tabular}{lr}
\textrm{Parameter f$_{sig}$} & \textrm{Value} \\
\hline
\\[-9pt]
decay rate $\Gamma$ & $(1.64\pm 0.95)\times10^{-24}$\,d$^{-1}$ \\
acceptance $\epsilon$ & $(98.3\pm0.1)\%$ \\
abundance $\eta$ & $(8.40\pm0.07)\times10^{-4}$ \\
exposure $mt$ & $(7636\pm45)$\,kg$\cdot$d \\
peak position $\mu_\mathrm{sig}$ & $(64.34\pm0.36)$\,keV \\
peak width $\sigma_\mathrm{sig}$ & $(4.08\pm0.26)$\,keV \\
background slope $a_{bkg}$ & $(0.103\pm0.016)$\,keV$^{-1}$ \\
background constant $c_{bkg}$ & $37.73\pm1.31$\, \\
\hline
\hline
\\[-9pt]
\textrm{Parameter f$_{bkg}$} & \textrm{Value} \\
\hline
\\[-9pt]
background slope $a_{bkg}$ & $(0.101\pm0.016)$\,keV$^{-1}$ \\
background constant $c_{bkg}$ & $38.22\pm1.29$\,\\
\end{tabular}
\end{ruledtabular}
\end{table}

A comparison of the current experimental half-life limits is shown in Table~\ref{tab:limits}. Since our analysis is more accurate it supersedes the previous limit given in Ref.~\cite{mei13} which made use of the publicly available XENON100 data. The larger mass of the XMASS experiment (835\,kg) results in better self-shielding capabilities and consequently a lower background, and thus makes the experiment more sensitive to 2$\nu$2K.
 
\begin{table}
\caption{\label{tab:limits} Current experimental limits (at 90\% confidence/credibility level) on the half-life of two-neutrino double K-capture (2$\nu$2K) of $^{124}$Xe. Our work supersedes the limit by Mei et al. \cite{mei13} which was based on publicly available XENON100 data.}
\begin{ruledtabular}
\begin{tabular}{ll}
\textrm{Reference} & $T_{1/2}$ ($10^{21}$\,yr) \\
\hline
Abe et al. (XMASS) \cite{abe15} & $>4.7$ \\
Gavrilyuk et al. \cite{gavrilyuk15} & $>2.0$ \\
Mei et al. \cite{mei13} & $>1.66$ \\
this work & $>0.65$ \\
\end{tabular}
\end{ruledtabular}
\end{table}
 
\begin{figure}
 \centering
 \includegraphics[width = 0.5\textwidth]{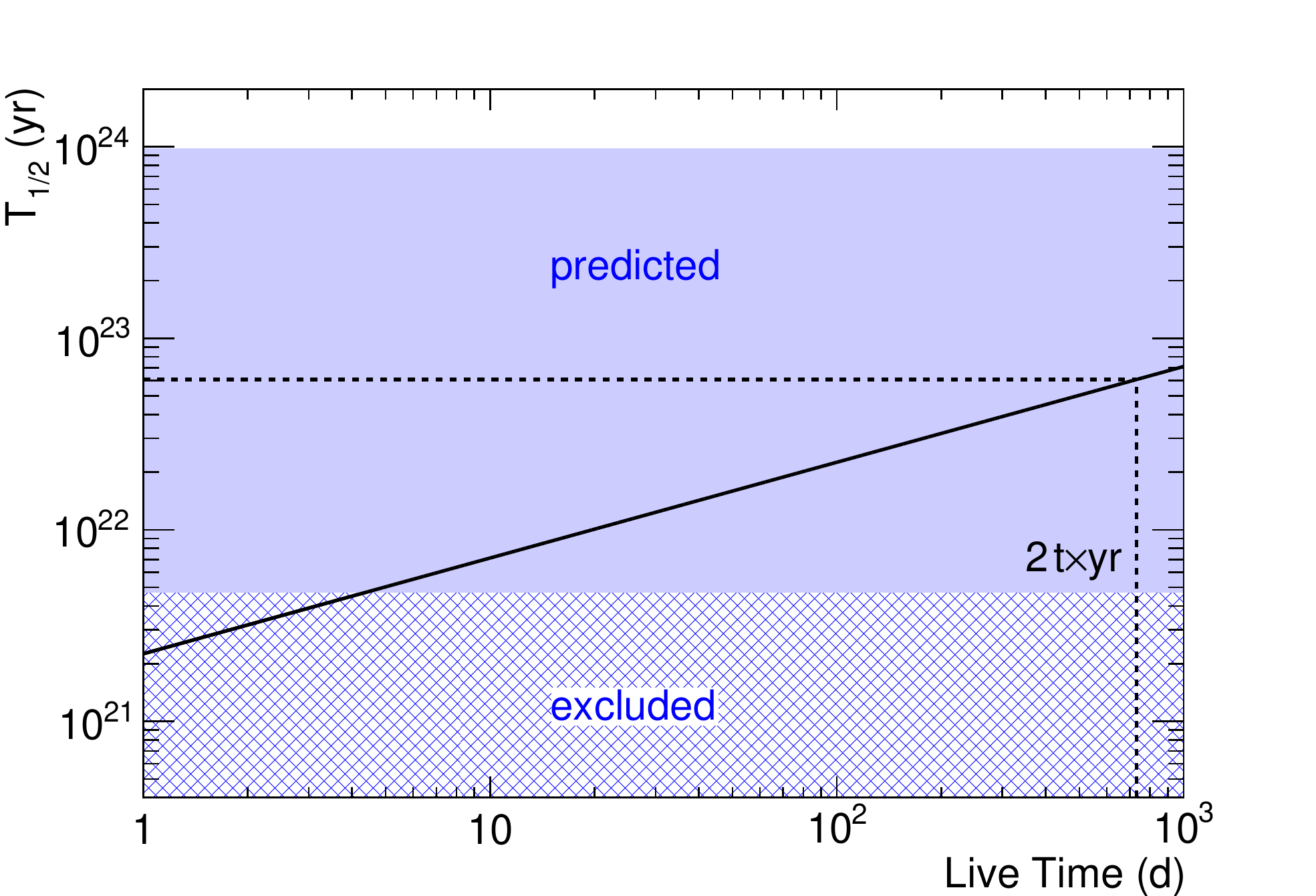}
 \caption{\label{fig:sensitivity} Expected sensitivity of XENON1T for 2$\nu$2K of $^{124}$Xe assuming a 1\,t fiducial volume. The blue-hatched area indicates the parameter space excluded by current experiments \cite{abe15}. The blue shaded area shows the range of predicted half-lives \cite{singh07}.}
 \end{figure}

The successor of XENON100, the XENON1T experiment \cite{aprile15}, is based on the same detector technology but with an increased total target mass of 2\,t and a reduced background. It is currently in the commissioning phase. The sensitivity of XENON1T for 2$\nu$2K of $^{124}$Xe was investigated using the expected background spectrum in a 1\,t fiducial volume \cite{aprile15}. We assumed the same energy resolution as in XENON100. This assumption is conservative as the energy resolution is related to the light yield, which is expected to be about a factor of two higher in XENON1T \cite{aprile15}.
We used the Bayesian approach for a simple counting experiment with known background to estimate the sensitivity on the half-life in XENON1T.
The likelihood is defined as
\begin{equation}
\mathcal{L} = \frac{(N_\mathrm{bkg}+N_\mathrm{sig})^{N_\mathrm{obs}}e^{-(N_\mathrm{bkg}+N_\mathrm{sig})}}{N_\mathrm{obs}!}\,,
\end{equation}
where $N_\mathrm{obs}=N_\mathrm{bkg}$ is the expected number of counts in the $\pm3\sigma$ region around the 2$\nu$2K peak, and $N_\mathrm{sig}$ is the number of signal counts. The first is a function of live-time with an expected value of 4.82 counts per day. For the latter a uniform prior was chosen.
The limit on the half-life $T_{1/2}$ was calculated as
\begin{equation}
T_{1/2}>\frac{\ln(2)\eta_\mathrm{nat}mt }{M_\mathrm{Xe}N_\mathrm{lim}}\,,
\end{equation}
where $\eta_\mathrm{nat}=9.52\times10^{-4}$ is the natural abundance of $^{124}$Xe and $N_\mathrm{lim}$ is the 90\% quantile of the posterior distribution for $N_\mathrm{sig}$.
The 90\% credibility limit on the half-life as a function of measurement time is shown in Fig.~\ref{fig:sensitivity} for a 1\,t fiducial target. With only five live days of XENON1T, we expect to reach a sensitivity exceeding the current best experimental limit. With an exposure of 2\,t$\cdot$yr we expect to reach a half-life limit of $6.1\times10^{22}$\,yr.
 
\section{\label{sec:level5}Conclusions}
 We have conducted a search for 2$\nu$2K of $^{124}$Xe using 7636\,kg$\cdot$d of XENON100 data. No significant signal was observed leading to a lower 90\% credibility limit on the half-life of $6.5\times10^{20}$\,yr. This result supersedes the previous limit of $>1.66\times10^{21}$\,yr \cite{mei13} from an external analysis of published XENON100 data. We have shown that the XENON1T experiment is expected to probe half-lives up to a value of $6.1\times10^{22}$\,yr after an exposure of 2\,t$\cdot$yr. Since the XENON1T detector was also designed to measure higher energy signals more accurately than XENON100 it offers the possibility to study neutrinoless double electron capture as well as electron capture with positron emission or double positron decay where the main part of the observable energy is above 1\,MeV \cite{barros14}. Moreover, future multi-ton target experiments such as XMASS-II\,\cite{abe15}, LZ\,\cite{mei13}, XENONnT\,\cite{aprile15} and DARWIN\,\cite{Darwin16} will have the sensitivity to investigate the parameter space even further.
 
\section*{\label{sec:level6}Acknowledgments}
We thank Dieter Frekers and Jouni Suhonen for useful discussion. We gratefully acknowledge support from: the National
Science Foundation, Swiss National Science Foundation,
Deutsche Forschungsgemeinschaft, Max Planck
Gesellschaft, Foundation for Fundamental Research on
Matter, Weizmann Institute of Science, I-CORE, Initial
Training Network Invisibles (Marie Curie Actions, PITNGA-2011-289442),
Fundacao para a Ciencia e a Tecnologia,
Region des Pays de la Loire, Knut and Alice Wallenberg
Foundation, and Istituto Nazionale di Fisica Nucleare.
We are grateful to Laboratori Nazionali del Gran
Sasso for hosting and supporting the XENON project.

\bibliography{Bibliography.bib}

\end{document}